\documentclass[manuscript=communication]{achemso}

\usepackage[version=3]{mhchem}
\usepackage[T1]{fontenc}
\usepackage[caption=false]{subfig}
\usepackage{soul}
\usepackage{color}

\usepackage{enumerate}
\usepackage{array}

\newcommand{\corr}[2]{#1}

\newcommand{\esi}{electronic supporting information}

\newcommand{\esis}{SI}
\newcommand{\titletext}{Identifying the Mechanism of Continued Growth of the Solid-Electrolyte Interphase}
\newcommand{\SOC}{SoC}
\newcommand{\SOCs}{SoCs}
\newcommand{\SOCv}{\mathrm{SoC}}

\newcommand{\jseil}{j_\mathrm{SEI}^{\mathrm{Li}_\mathrm{I}}}

\newcommand{\Qirr}{Q_\mathrm{irr}}
\newcommand{\Qirrlin}{Q_\mathrm{irr}^\mathrm{lin}}
\newcommand{\Qirrsei}{Q_\mathrm{irr}^\mathrm{SEI}}
\newcommand{\Qirrseiz}{Q_\mathrm{irr,0}^\mathrm{SEI}}

\newcommand{\clii}{c_{\mathrm{Li}_\mathrm{I}}}
\newcommand{\cliiz}{c_{\mathrm{Li}_\mathrm{I},0}}
\newcommand{\cliimax}{c_{\mathrm{Li}_\mathrm{I},\mathrm{max}}}
\newcommand{\lii}{\mathrm{Li}_\mathrm{I}}

\newcommand{\li}{\mathrm{Li}}

\newcommand{\dt}{\partial_t}

\newcommand{\DLi}{D_{\mathrm{Li}_\mathrm{I}}}
\newcommand{\DEC}{D_\mathrm{EC}}
\newcommand{\cEC}{c_\mathrm{EC}}

\newcommand{\dC}{^\mathrm{o}\mathrm{C}}

\newcommand{\vslilip}{vs. Li/Li$^+$}
\newcommand{\mulixc}{\mu^{\li}_{\mathrm{Li}_x\mathrm{C}_6}}
\newcommand{\musei}{\mu^{\lii}_{\mathrm{SEI}}}
\newcommand{\museiz}{\mu^{\lii}_{\mathrm{SEI},0}}

\newcolumntype{L}[1]{>{\raggedright\let\newline\\\arraybackslash\hspace{0pt}}m{#1}}
\newcolumntype{C}[1]{>{\centering\let\newline\\\arraybackslash\hspace{0pt}}m{#1}}
\newcolumntype{R}[1]{>{\raggedleft\let\newline\\\arraybackslash\hspace{0pt}}m{#1}}

\usepackage[nokeyprefix]{refstyle}
\usepackage{xr-hyper}
\usepackage{hyperref}
\usepackage{cleveref}
\usepackage{xcolor}
\hypersetup{
    colorlinks,
    linkcolor={blue!70!black},
    filecolor={blue!70!black},
    citecolor={blue!70!black},
    urlcolor={blue!70!black}
}

\author{Fabian Single}
\affiliation[German Aerospace Center]{German Aerospace Center (DLR), Institute of Engineering Thermodynamics, Pfaffenwaldring 38-40, 70569 Stuttgart, Germany}
\alsoaffiliation[Helmholtz Institute Ulm]{Helmholtz Institute Ulm (HIU), Helmholtzstra\ss e, 11, 89081 Ulm, Germany}
\author{Arnulf Latz}
\alsoaffiliation[German Aerospace Center]{German Aerospace Center (DLR), Institute of Engineering Thermodynamics, Pfaffenwaldring 38-40, 70569 Stuttgart, Germany}
\alsoaffiliation[Helmholtz Institute Ulm]{Helmholtz Institute Ulm (HIU), Helmholtzstra\ss e, 11, 89081 Ulm, Germany}
\affiliation[University of Ulm]{University of Ulm, Institute of Electrochemistry, Albert-Einstein-Allee 47, 89069 Ulm, Germany}
\author{Birger Horstmann}
\affiliation[German Aerospace Center]{German Aerospace Center (DLR), Institute of Engineering Thermodynamics, Pfaffenwaldring 38-40, 70569 Stuttgart, Germany}
\alsoaffiliation[Helmholtz Institute Ulm]{Helmholtz Institute Ulm (HIU), Helmholtzstra\ss e, 11, 89081 Ulm, Germany}
\email{birger.horstmann@dlr.de}

\title{\titletext}

\abbreviations{SEI}
\keywords{SEI,Lithium-Ion,Battery}

\begin{document}

\setlength{\fboxrule}{0 pt}

\begin{abstract}
Continued growth of the solid electrolyte interphase (SEI) is the major reason for capacity fade in modern lithium-ion batteries.
This growth is made possible by a yet unidentified transport mechanism that limits the passivating ability of the SEI towards electrolyte reduction.
We, for the first time, differentiate the proposed mechanisms by analyzing their dependence on the electrode potential.
Our calculations are compared to recent experimental capacity fade data.
We show that the potential dependence of SEI growth facilitated by solvent diffusion, electron conduction, or electron tunneling qualitatively disagrees with the experimental observations.
Only diffusion of Li-interstitials results in a potential dependence matching to the experiments.
Therefore, we identify Li-interstitial diffusion as the cause of long-term SEI growth.
\end{abstract}

\begin{figure}[t!]
\subfloat[\label{fig:toca}]{}
\subfloat[\label{fig:tocb}]{}
\subfloat[\label{fig:tocc}]{}
\subfloat[\label{fig:tocd}]{}
\includegraphics[width=\columnwidth]{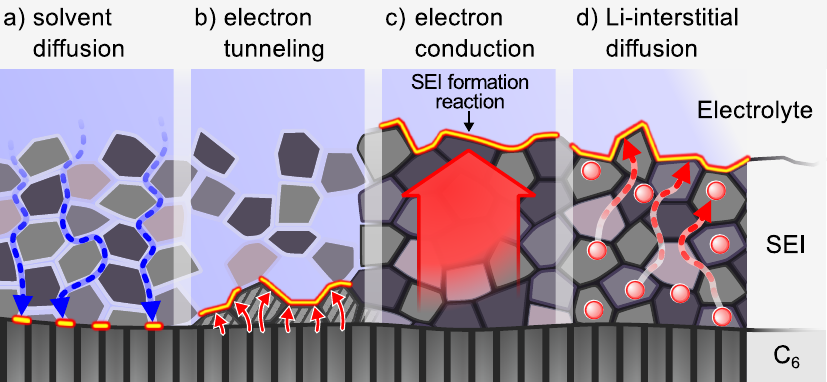}
\caption{\label{fig:toc} Schematic of four different transport mechanisms which have been suggested to cause long-term SEI growth.
a) Solvent diffusion through small SEI pores.
b) Electron tunneling through a thin and dense inner SEI layer.
c) Electron conduction through the SEI.
d) Diffusion of neutral Li-interstitials through the SEI.
The SEI formation reaction takes place at different interfaces depending on the mechanism, marked yellow/red.}
\end{figure}

\phantomsection
\addcontentsline{toc}{section}{Introduction}
Despite all recent advances, lithium-ion batteries still suffer from continued capacity fade which ultimately limits battery lifetime.
A multitude of processes contribute to the capacity fade.
These mechanisms depend on operating conditions as well as on battery chemistry.
However, generally, anodic side reactions are found to be the main contributor to capacity fade \cite{Keil2016,Keil2017}.
These reactions reduce electrolyte components, e.g., ethylene dicarbonate \corr{(EC)}{}, while irreversibly consuming cyclable lithium.
They proceed rapidly on a pristine electrode until they are suppressed by the solid electrolyte interphase (SEI).
SEI is a thin film which covers the electrode surface and consists of insoluble products of anodic reactions \cite{peled1979electrochemical,verma2010review,Kang2008,Xiao2009,Nie2013,Nie2013-2}.

Atomistic simulation methods cover the short-term SEI formation occurring during the first few battery cycles (see Bedrov et al. \cite{bedrov2017li+}).
After this formation stage, the long-term SEI growth rate is limited by the rate at which SEI precursors traverse the SEI.
The transport mechanism enabling this flux is referred to as the long-term growth mechanism (LTGM).
Even though numerous publications discuss long-term SEI growth
\cite{Ploehn2004,Pinson2012,Tang2012,
Li2015,
Broussely2001,Christensen2004,Single2016,Single2017,roder2017multi,
colclasure2011modeling,roder2016multi,hao2017mesoscale}
, the LTGM has not been identified.
Several different LTGMs are suggested and studied with continuum models as depicted in \cref{fig:toc}.
\begin{enumerate}[a)]
\item \label{mech1} Diffusion of solvent/salt molecules/anions through nano-sized SEI pores \cite{Ploehn2004,Pinson2012,Tang2012,Single2016,Single2017}.
\item \label{mech2} Electron tunneling through a dense, inner layer of the SEI \cite{Tang2012,Li2015}.
\item \label{mech3} Electron conduction through the SEI \cite{Broussely2001,Christensen2004,Tang2012,Single2016,Single2017,roder2017multi}.
\item \label{mech4} Diffusion of neutral lithium interstitials ($\lii$) \cite{Shi2012,Single2017}.
\end{enumerate}
Importantly, these four mechanisms predict a similar evolution of long-term capacity fade.
Besides electron tunneling, all mechanisms directly result in the experimentally observed $\sqrt{t}$ dependence of capacity fade.
Electron tunneling predicts a $\ln{t}$ dependence which fits reasonably well to the $\sqrt{t}$ behaviour if another contribution linear in time is added \cite{Broussely2001,Smith2011}.
Such a term can be attributed to multiple processes which we discuss below.

Therefore, additional dependencies must be studied in order to identify the correct mechanism (or to rule out others).
For this reason, we have introduced a SEI model predicting SEI morphology in previous studies \cite{Single2016,Single2017}.
If SEI porosity and thickness is measured, e.g., with neutron reflectometry \cite{steinhauer2017situ}, our model offers an alternative feature to compare and validate SEI theory with experiments.
Furthermore, we used our models to compare different LTGMs and their response to small porosity fluctuations.
Based on this comparison, we concluded that solvent diffusion is unlikely to be LTGM \cite{Single2017}.

In this paper, we finally identify the LTGM by comparing the rate of SEI formation at different electrode potentials to experimental capacity fade data.
This dependence has already been used by Tang et al.\cite{Tang2012} to rule out solvent diffusion as possible LTGM.
We, however, perform a more comprehensive comparison with more recent experimental data provided by Keil et al. \cite{Keil2016,Keil2017}
 
The capacity fade of commercial NCA cells has been measured during long-term open circuit storage. \cite{Keil2016,Keil2017}
Individual cells were stored at one of 16 different states of charge ({\SOC}), each corresponding to a specific anode potential.
These cells were stored for 9.5 months at 50$\dC$.
As capacity fade during open circuit storage leads to self-discharge, checkup sequences were regularly performed.
After these sequences, the {\SOC} referenced to the current cell capacity reached its initial value.
The complete measurements are presented in fig. 8 of ref. \citenum{Keil2017}.
They provide a unique opportunity to compare \emph{all} plausible LTGMs with respect to the emerging potential dependence.

\begin{figure}[ht!]
\subfloat[\label{fig:lrc}]{}
\subfloat[\label{fig:socu}]{}
\includegraphics{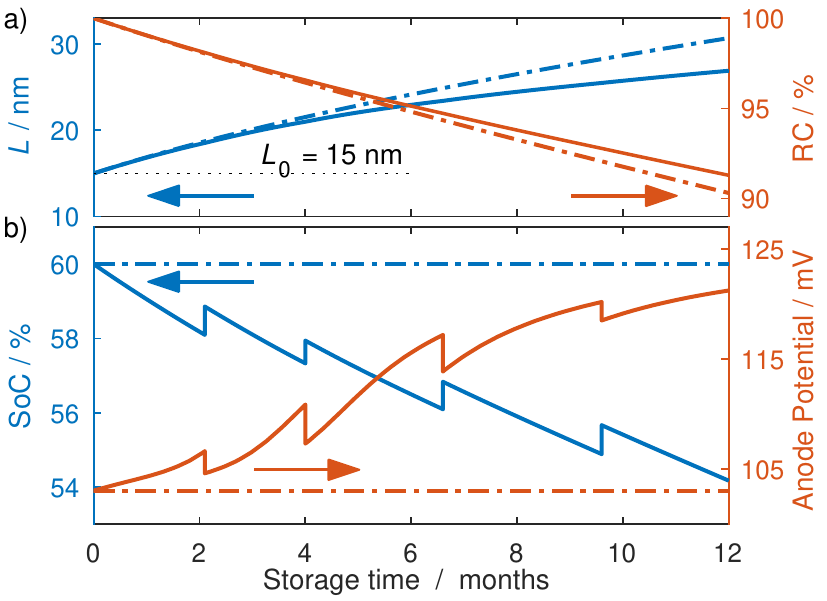}
\caption{Simulation of a storage experiment with 60\% initial {\SOC} and four checkup cycles ($\lii$-diffusion as the LTGM).
a) Evolution of SEI thickness $L$ and the relative capacity of the cell.
b) Evolution of the {\SOC} referenced to the original cell capacity and $U$, the corresponding anode potential \vslilip.
\corr{Jumps in {\SOC} and $U$ correspond to checkup sequences which were part of the experimental procedure.}{}
The dash-dotted lines show the evolution of these quantities if the {\SOC}/electrode potential is assumed constant.}
\label{fig:fcsim}
\end{figure}

\begin{figure}[h!]
\subfloat[\label{fig:ocv}]{}
\subfloat[\label{fig:res1}]{}
\includegraphics{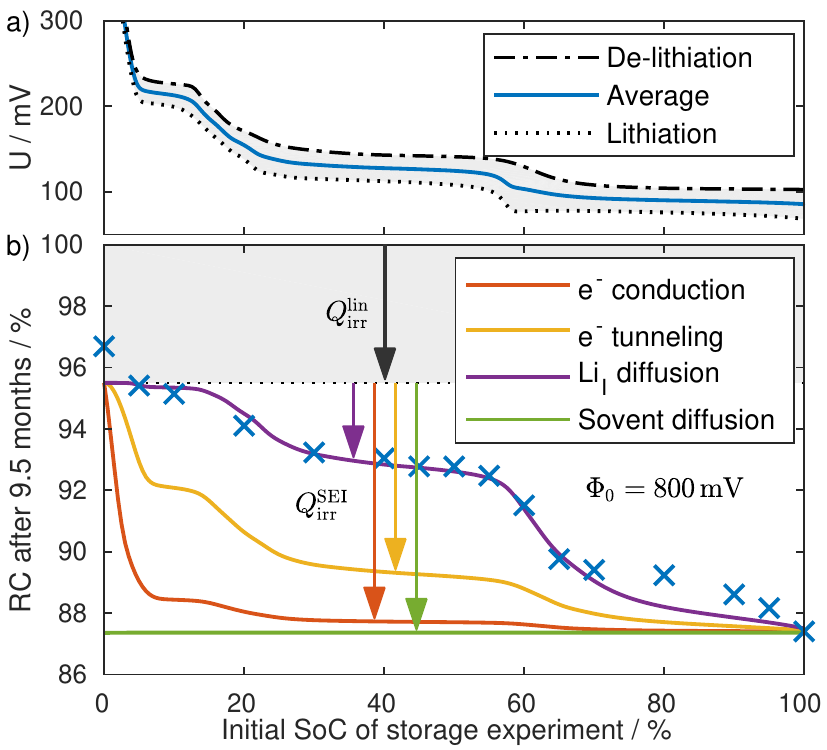}
\caption{\label{fig:res}
a) Open circuit voltage of the negative electrode gained by averaging the lithiation and delithiation voltages (half cell, cycled at C/20).
b) Experimentally obtained relative capacity after 9.5 months of storage (crosses) compared to that predicted by four different LTGMs (lines).}
\end{figure}

\phantomsection
\addcontentsline{toc}{section}{Theory}
To this aim, we formulate a model and simulate the evolution of the irreversible capacity $\Qirr$ and the {\SOC} of a single battery during the experiment.
The full model is derived in the {\esi} (\esis).
An example simulation is presented in \cref{fig:fcsim}.
During storage, the SEI thickness increases while the relative capacity (RC) of the cell decreases, see \cref{fig:lrc}.
In \cref{fig:socu} we show the evolution of the {\SOC} which decreases smoothly.
Note that in this paper we reference the {\SOC} to the capacity of a fresh cell.
With this definition, the {\SOC} increases during the periodically performed check-up sequences, but does not reach its initial value. 
Also shown is the corresponding potential of the negative electrode $U$.
We determine this potential for a given {\SOC} with the OCV measured by Keil et al. \cite{Keil2016}, shown in \cref{fig:ocv}.

In our model we assume that capacity fade is the sum of two distinct contributions, $\Qirr = \Qirrsei + \Qirrlin$.
The former part $\Qirrsei$ includes the amount of lithium that is irreversibly consumed by SEI formation during the storage experiment.
It is directly coupled to the SEI thickness and the rate $\partial_t \Qirrsei$ can have a strong dependence on the anode potential, depending on the LTGM assumed.
For the latter part $\Qirrlin$ we assume no such dependence.
This contribution is assumed to increase at a constant rate in time and factors in various mechanisms, e.g., rapid reformation of SEI due to cracks and delamination of the existing film.
Cracks and delamination of SEI occur during the periodic checkup sequences.
Such physical stress also causes electrode particles to loose contact to the current collector which causes an irreversible loss of lithium.

\phantomsection
\addcontentsline{toc}{section}{Results}
We now simulate the storage experiment using different initial {\SOCs} with each of the SEI formation mechanism mentioned above.
The capacity fade from these simulations is compared to the experimentally measured one in \cref{fig:res1}.
The {\SOC} dependence of the relative capacity is evident and a correlation to the potential of the negative electrode (shown above) can be made out with the naked eye.
Capacity fade experiences a significant increase at {\SOCs} larger than 60\% which correlates to the potential step of the OCV.
Furthermore, capacity fade remains nearly constant in {\SOC} regions which correspond to the voltage plateaus of graphite.

As elaborated above, we split capacity fade into two contributions.
During storage every cell looses the same amount of charge to processes summarized in $\Qirrlin$.
This contribution is  independent of the {\SOC} and serves as a baseline for the relative capacity in \cref{fig:res1} (dotted line).
In addition, $\Qirrsei$, is lost to continued SEI formation.
This contribution depends on the RTLM assumed and features a {\SOC} dependence.

It is evident that SEI formation facilitated by solvent diffusion does not depend on the potential and cannot reproduce the experimental data.
Both, electron conduction and electron tunneling lead to a potential dependence, but it does not agree with the experiment.
These mechanisms fail to reproduce the pronounced change of the relative capacity at 60\% {\SOC}.
Instead, they predict a high potential sensitivity at {\SOCs} between 0 and 20\%.

$\lii$ diffusion is the only LTGM that predicts capacity fade in excellent agreement with the experiment.
This agreement is due to the exponential dependence of capacity fade on electrode potential, see \cref{eq:sold}.
$\lii$ diffusion correctly describes the capacity fade increase between 10 and 30\% {\SOC} as well as the one between 50 and 70\% {\SOC}.
Small deviations between this model and the experiment are only present at zero {\SOC} and at high {\SOCs}.

We attribute the deviation at zero {\SOC} to the mismatch in electrode areas.
Because the coated anode area is larger than the coated cathode area, an overhang area of the anode has no opposed cathode counterpart.
The overhang anode acts as a lithium reservoir at small {\SOCs}. 
We expect a capacity increase of $\sim$1\% due to overhang anode at zero {\SOC} as elaborated in the \esis.
Taking this into account results in a good agreement with the measured capacity at zero {\SOC}.
We attribute the high {\SOC} mismatch to two effects.
Because the overhang anode area accumulates lithium during battery storage at low anode potentials, cell capacity is reduced.
Most importantly, high {\SOC}s correspond to high cathode potentials which enable electrolyte oxidation reactions.
These reactions increase the amount of cyclable lithium in the cell \cite{deshpande2015limited}.
Modeling these partially counteracting effects is beyond the scope of this work.
Therefore, small deviations between our model and the experiment are to be expected at high {\SOCs}.

Now, we evaluate if alternative parameter choices can improve the agreement between electron conduction / electron tunneling and the experiment shown in \cref{fig:res1}.
The first option is to assign a potential dependence to one essential model parameter.
For instance, the electron conductivity $\kappa$ for electron conduction or the parameter $\delta$ for $e^-$ tunneling, see Li et al. \cite{Li2015}. 
However, this seems highly speculative if no physical explanation is given.
We choose another approach.
Both mechanisms are highly sensitive to the {\SOC} if the corresponding anode potential is close to $\Phi_0$, the onset potential of SEI formation.
Therefore, it is possible to increase the sensitivity at 60\% {\SOC} by lowering $\Phi_0$.
We explore this option in the \esis, showing that, theoretically, both LTGMs are able to predict the pronounced relative capacity drop at 60\%, see \cref{A-fig:res2}.
However, in turn, they no longer predict a capacity drop at  30\% {\SOC}.
Most importantly, the required values for $\Phi_0$ are far below any value reported in literature \cite{Xiao2009,Borodin2015}.

\phantomsection
\addcontentsline{toc}{section}{Summary and Conclusion}
To conclude, we compare SEI growth based on four long‐term growth mechanisms (LTGM) to an experimental study.
Only $\lii$ diffusion results in a promising agreement with the experiment which makes it a very likely candidate for the LTGM.
Solvent diffusion does not reproduce a {\SOC} dependence and is very unlikely to be the LTGM.
Both, electron conduction and electron tunneling predict a {\SOC} dependence but it does not agree with the experiment for any reasonable choice of parameters.

\noindent\rule{\columnwidth}{1pt}\\

\phantomsection
\addcontentsline{toc}{section}{Computational}
\textbf{Computational Methods}
\corr{We now derive a simplified  capacity fade model which applies to neutral $\lii$ diffusing towards the SEI/electrolyte interface.
Li-ions take up an electron at the electrode and release it at the SEI/electrolyte interface for electrolyte reduction.
The following derivation applies to alternative carries of negative charge as well.}{}
The full model used to create the data in \cref{fig:res1} is presented in the \esis.
We assume that the SEI is a homogeneous film that spans from $x=0$ (electrode/SEI interface) to $x=L$ (SEI/electrolyte interface).
SEI thickness $L$ is directly related to $\Qirrsei$ via
\begin{align}
\label{eq:lq}
L = \frac{V}{s} \frac{\Qirrsei}{AF} + L_0,
\end{align}
where $V$ is the mean partial molar volume of the SEI and $s$ is the mean stoichiometric coefficient of $\lii$ in the SEI formation reaction.
$L_0$ is the SEI thickness at the start of the experiment and $A$ is the surface area of the negative electrode.

The $\lii$ diffusion flux in the SEI is given by Fick's law (in units of A/m$^2$),
\begin{align}
\label{eq:dif}
\jseil 	&= - F \DLi \cdot \nabla \clii \notag \\
	&\approx - F \DLi \cdot \frac{\clii|_{x=L} - \clii|_{x=0}}{L} .
\end{align}
The approximation in the second line is possible because SEI is homogeneous and reactions take place at the SEI/electrolyte interface only.
This is also true if the SEI has nano-sized pores, as we have shown in previous works \cite{Single2016,Single2017}.
\corr{Note that we do not specify the diffusion pathway.
Interstitials could diffuse through the bulk SEI, pass through a selected SEI compound, or move along nano-sized SEI pores.}{}
The $\lii$ flux directly induces capacity fade, i.e., $\dt \Qirrsei = A \jseil$.
Together with \cref{eq:lq} and \cref{eq:dif}, this yields a differential equation for $\Qirrsei$
\begin{align}
\label{eq:dqlii}
\dt\Qirrsei = \frac{A^2 sF^2D}{V} \frac{\clii|_{x=0} - \clii|_{x=L}}{\Qirrsei + \Qirrseiz},
\end{align}
where $\Qirrseiz =  s A L_0 F / V$ is the capacity corresponding to $L_0$.

Next, we determine the $\lii$ concentration at $x=0$ and $x=L$.
At the electrode/SEI interface, interstitials are injected into the SEI.
We assume that injection is a fast process and that graphite is in thermodynamic equilibrium with the SEI across the interface
\begin{align}
\label{eq:mueqmu}
\mulixc &= \musei \notag \\
&= \museiz + RT \ln \frac{\clii|_{x=0}}{\cliimax}.
\end{align}
$\cliimax$ is the maximal interstitial concentration.
$\mu^{\lii}_{\mathrm{SEI},0}$ is constant and can be determined with DFT methods.
This has been done by Shi et  al. \cite{Shi2012} for a Li$_2$CO$_3$ host lattice.
The electrochemical potential of Li in the electrode is equal to $-FU$.\cite{Latz2011,Latz2013}
Thus, we can express the interstitial concentration at the interface as 
\begin{align}
\label{eq:clii}
\clii|_{x=0} = \cliiz \cdot \exp{\bigg(-\frac{FU}{RT} \bigg)},
\end{align}
where $\cliiz$ is the interstitial concentration at $U=0\,$V.
$\cliiz$ is a model parameter and absorbs all constant contributions in \cref{eq:mueqmu}.
At the SEI/electrolyte interface lithium interstitials do not accumulate.
Instead, they are consumed by the fast SEI formation reaction, i.e., $\clii|_{x=L}=0$. 
This is the assumption of transport limited growth.

\Cref{eq:dqlii} can be solved analytically in the simplified case of constant electrode potential. 
We illustrate the accuracy of this assumption in \cref{fig:fcsim}.
Note that we solve our full model numerically without it.
Similar equations are derived for electron conduction and solvent diffusion in \cref{A-eq:dq,A-eq:flux} of the \esis.
The corresponding solutions in the order solvent diffusion, $e^-$ tunneling, $e^-$ conduction, and $\lii$ diffusion are
\begin{subequations}
\label{eq:sol}
\begin{align}
\Qirrsei &= A \Gamma \sqrt{F\DEC\cEC} 					\cdot \sqrt{t+\tau} -\Qirrseiz	     		,\label{eq:sola}\\
\Qirrsei &= A \cdot \alpha(\SOCv) \cdot \ln\big[1 + \beta(\SOCv)\cdot t \big]			     		,\label{eq:solb}\\
\Qirrsei &= A \Gamma 			\sqrt{\kappa\big(\Phi_0 - U(\SOCv)\big)}	\cdot \sqrt{t+\tau}-\Qirrseiz  	,\label{eq:solc}\\
\Qirrsei &= A \Gamma \sqrt{F\DLi\cliiz 	e^{-\frac{FU(\SOCv)}{RT}}} 	\cdot \sqrt{t+\tau}-\Qirrseiz	     	.\label{eq:sold} 	 
\end{align}
\end{subequations}
Here, $\Gamma$ equals $\sqrt{2sF/V}$ and $\tau$ is determined by the initial SEI thickness $L_0(\Qirrseiz)$ through the requirement $\Qirrsei(t=0)=0$.
\Cref{eq:solb} is the electron tunneling model derived by Li et al., see eq. 30 in ref. \citenum{Li2015}.
We list the model parameters in \cref{A-tbl:param} in the \esis \cite{borodin2006molecular,zhuang2005lithium,edstrom2006new,kulova2006temperature}.

We want to highlight another way in which electron tunneling differs from the other LTGMs.
It is the only mechanism for which time dependence and {\SOC} dependence cannot be separated.
This means that $\Qirrsei$ cannot be written in the form $f(\SOCv)\cdot g(t)$, see \cref{eq:sol}.
Therefore, for electron tunneling, the qualitative shape of the predicted RC in \cref{fig:res1} depends on the time it is evaluated at.
This behavior is not observed in the experiment \cite{Keil2016}.

\begin{acknowledgement}
We thank Peter Keil for providing us with details and data of his experiments.
This work is supported by the German Federal Ministry of Education and Research (BMBF) in the project Li-EcoSafe (03X4636A).
\end{acknowledgement}

\bibliography{references}

\providecommand{\latin}[1]{#1}
\providecommand*\mcitethebibliography{\thebibliography}
\csname @ifundefined\endcsname{endmcitethebibliography}
  {\let\endmcitethebibliography\endthebibliography}{}
\begin{mcitethebibliography}{33}
\providecommand*\natexlab[1]{#1}
\providecommand*\mciteSetBstSublistMode[1]{}
\providecommand*\mciteSetBstMaxWidthForm[2]{}
\providecommand*\mciteBstWouldAddEndPuncttrue
  {\def\EndOfBibitem{\unskip.}}
\providecommand*\mciteBstWouldAddEndPunctfalse
  {\let\EndOfBibitem\relax}
\providecommand*\mciteSetBstMidEndSepPunct[3]{}
\providecommand*\mciteSetBstSublistLabelBeginEnd[3]{}
\providecommand*\EndOfBibitem{}
\mciteSetBstSublistMode{f}
\mciteSetBstMaxWidthForm{subitem}{(\alph{mcitesubitemcount})}
\mciteSetBstSublistLabelBeginEnd
  {\mcitemaxwidthsubitemform\space}
  {\relax}
  {\relax}

\bibitem[Keil \latin{et~al.}(2016)Keil, Schuster, Wilhelm, Travi, Hauser, Karl,
  and Jossen]{Keil2016}
Keil,~P.; Schuster,~S.~F.; Wilhelm,~J.; Travi,~J.; Hauser,~A.; Karl,~R.~C.;
  Jossen,~A. \emph{Journal of The Electrochemical Society} \textbf{2016},
  \emph{163}, A1872--A1880\relax
\mciteBstWouldAddEndPuncttrue
\mciteSetBstMidEndSepPunct{\mcitedefaultmidpunct}
{\mcitedefaultendpunct}{\mcitedefaultseppunct}\relax
\EndOfBibitem
\bibitem[Keil and Jossen(2017)Keil, and Jossen]{Keil2017}
Keil,~P.; Jossen,~A. \emph{Journal of The Electrochemical Society}
  \textbf{2017}, \emph{164}, 6066--6074\relax
\mciteBstWouldAddEndPuncttrue
\mciteSetBstMidEndSepPunct{\mcitedefaultmidpunct}
{\mcitedefaultendpunct}{\mcitedefaultseppunct}\relax
\EndOfBibitem
\bibitem[Peled(1979)]{peled1979electrochemical}
Peled,~E. \emph{Journal of The Electrochemical Society} \textbf{1979},
  \emph{126}, 2047--2051\relax
\mciteBstWouldAddEndPuncttrue
\mciteSetBstMidEndSepPunct{\mcitedefaultmidpunct}
{\mcitedefaultendpunct}{\mcitedefaultseppunct}\relax
\EndOfBibitem
\bibitem[Verma \latin{et~al.}(2010)Verma, Maire, and
  Nov{\'a}k]{verma2010review}
Verma,~P.; Maire,~P.; Nov{\'a}k,~P. \emph{Electrochimica Acta} \textbf{2010},
  \emph{55}, 6332--6341\relax
\mciteBstWouldAddEndPuncttrue
\mciteSetBstMidEndSepPunct{\mcitedefaultmidpunct}
{\mcitedefaultendpunct}{\mcitedefaultseppunct}\relax
\EndOfBibitem
\bibitem[Kang \latin{et~al.}(2008)Kang, Abraham, Xiao, and Lucht]{Kang2008}
Kang,~S.~H.; Abraham,~D.~P.; Xiao,~a.; Lucht,~B.~L. \emph{Journal of Power
  Sources} \textbf{2008}, \emph{175}, 526--532\relax
\mciteBstWouldAddEndPuncttrue
\mciteSetBstMidEndSepPunct{\mcitedefaultmidpunct}
{\mcitedefaultendpunct}{\mcitedefaultseppunct}\relax
\EndOfBibitem
\bibitem[Xiao \latin{et~al.}(2009)Xiao, Yang, Lucht, Kang, and
  Abraham]{Xiao2009}
Xiao,~A.; Yang,~L.; Lucht,~B.~L.; Kang,~S.-H.; Abraham,~D.~P. \emph{Journal of
  The Electrochemical Society} \textbf{2009}, \emph{156}, A318\relax
\mciteBstWouldAddEndPuncttrue
\mciteSetBstMidEndSepPunct{\mcitedefaultmidpunct}
{\mcitedefaultendpunct}{\mcitedefaultseppunct}\relax
\EndOfBibitem
\bibitem[Nie \latin{et~al.}(2013)Nie, Chalasani, Abraham, Chen, Bose, and
  Lucht]{Nie2013}
Nie,~M.; Chalasani,~D.; Abraham,~D.~P.; Chen,~Y.; Bose,~A.; Lucht,~B.~L.
  \emph{Journal of Physical Chemistry C} \textbf{2013}, \emph{117},
  1257--1267\relax
\mciteBstWouldAddEndPuncttrue
\mciteSetBstMidEndSepPunct{\mcitedefaultmidpunct}
{\mcitedefaultendpunct}{\mcitedefaultseppunct}\relax
\EndOfBibitem
\bibitem[Nie \latin{et~al.}(2013)Nie, Abraham, Chen, Bose, and
  Lucht]{Nie2013-2}
Nie,~M.; Abraham,~D.~P.; Chen,~Y.; Bose,~A.; Lucht,~B.~L. \emph{The Journal of
  Physical Chemistry C} \textbf{2013}, \emph{117}, 13403--13412\relax
\mciteBstWouldAddEndPuncttrue
\mciteSetBstMidEndSepPunct{\mcitedefaultmidpunct}
{\mcitedefaultendpunct}{\mcitedefaultseppunct}\relax
\EndOfBibitem
\bibitem[Bedrov \latin{et~al.}(2017)Bedrov, Borodin, and Hooper]{bedrov2017li+}
Bedrov,~D.; Borodin,~O.; Hooper,~J.~B. \emph{The Journal of Physical Chemistry
  C} \textbf{2017}, \emph{121}, 16098--16109\relax
\mciteBstWouldAddEndPuncttrue
\mciteSetBstMidEndSepPunct{\mcitedefaultmidpunct}
{\mcitedefaultendpunct}{\mcitedefaultseppunct}\relax
\EndOfBibitem
\bibitem[Ploehn \latin{et~al.}(2004)Ploehn, Ramadass, and White]{Ploehn2004}
Ploehn,~H.~J.; Ramadass,~P.; White,~R.~E. \emph{Journal of The Electrochemical
  Society} \textbf{2004}, \emph{151}, A456\relax
\mciteBstWouldAddEndPuncttrue
\mciteSetBstMidEndSepPunct{\mcitedefaultmidpunct}
{\mcitedefaultendpunct}{\mcitedefaultseppunct}\relax
\EndOfBibitem
\bibitem[Pinson and Bazant(2012)Pinson, and Bazant]{Pinson2012}
Pinson,~M.~B.; Bazant,~M.~Z. \emph{Journal of the Electrochemical Society}
  \textbf{2012}, \emph{160}, A243--A250\relax
\mciteBstWouldAddEndPuncttrue
\mciteSetBstMidEndSepPunct{\mcitedefaultmidpunct}
{\mcitedefaultendpunct}{\mcitedefaultseppunct}\relax
\EndOfBibitem
\bibitem[Tang \latin{et~al.}(2012)Tang, Lu, and Newman]{Tang2012}
Tang,~M.; Lu,~S.; Newman,~J. \emph{Journal of The Electrochemical Society}
  \textbf{2012}, \emph{159}, A1775--A1785\relax
\mciteBstWouldAddEndPuncttrue
\mciteSetBstMidEndSepPunct{\mcitedefaultmidpunct}
{\mcitedefaultendpunct}{\mcitedefaultseppunct}\relax
\EndOfBibitem
\bibitem[Li \latin{et~al.}(2015)Li, Danilov, Zhang, Chen, Yang, and
  Notten]{Li2015}
Li,~D.; Danilov,~D.; Zhang,~Z.; Chen,~H.; Yang,~Y.; Notten,~P. H.~L.
  \emph{Journal of the Electrochemical Society} \textbf{2015}, \emph{162},
  A858--A869\relax
\mciteBstWouldAddEndPuncttrue
\mciteSetBstMidEndSepPunct{\mcitedefaultmidpunct}
{\mcitedefaultendpunct}{\mcitedefaultseppunct}\relax
\EndOfBibitem
\bibitem[Broussely \latin{et~al.}(2001)Broussely, Herreyre, Biensan, Kasztejna,
  Nechev, and Staniewicz]{Broussely2001}
Broussely,~M.; Herreyre,~S.; Biensan,~P.; Kasztejna,~P.; Nechev,~K.;
  Staniewicz,~R.~J. \emph{Journal of Power Sources} \textbf{2001},
  \emph{97-98}, 13--21\relax
\mciteBstWouldAddEndPuncttrue
\mciteSetBstMidEndSepPunct{\mcitedefaultmidpunct}
{\mcitedefaultendpunct}{\mcitedefaultseppunct}\relax
\EndOfBibitem
\bibitem[Christensen and Newman(2004)Christensen, and Newman]{Christensen2004}
Christensen,~J.; Newman,~J. \emph{Journal of The Electrochemical Society}
  \textbf{2004}, \emph{151}, A1977\relax
\mciteBstWouldAddEndPuncttrue
\mciteSetBstMidEndSepPunct{\mcitedefaultmidpunct}
{\mcitedefaultendpunct}{\mcitedefaultseppunct}\relax
\EndOfBibitem
\bibitem[Single \latin{et~al.}(2016)Single, Horstmann, and Latz]{Single2016}
Single,~F.; Horstmann,~B.; Latz,~A. \emph{Phys. Chem. Chem. Phys.}
  \textbf{2016}, \emph{18}, 17810--17814\relax
\mciteBstWouldAddEndPuncttrue
\mciteSetBstMidEndSepPunct{\mcitedefaultmidpunct}
{\mcitedefaultendpunct}{\mcitedefaultseppunct}\relax
\EndOfBibitem
\bibitem[Single \latin{et~al.}(2017)Single, Horstmann, and Latz]{Single2017}
Single,~F.; Horstmann,~B.; Latz,~A. \emph{Journal of The Electrochemical
  Society} \textbf{2017}, \emph{164}, E3132--E3145\relax
\mciteBstWouldAddEndPuncttrue
\mciteSetBstMidEndSepPunct{\mcitedefaultmidpunct}
{\mcitedefaultendpunct}{\mcitedefaultseppunct}\relax
\EndOfBibitem
\bibitem[R{\"o}der \latin{et~al.}(2017)R{\"o}der, Braatz, and
  Krewer]{roder2017multi}
R{\"o}der,~F.; Braatz,~R.~D.; Krewer,~U. \emph{Journal of The Electrochemical
  Society} \textbf{2017}, \emph{164}, E3335--E3344\relax
\mciteBstWouldAddEndPuncttrue
\mciteSetBstMidEndSepPunct{\mcitedefaultmidpunct}
{\mcitedefaultendpunct}{\mcitedefaultseppunct}\relax
\EndOfBibitem
\bibitem[Colclasure \latin{et~al.}(2011)Colclasure, Smith, and
  Kee]{colclasure2011modeling}
Colclasure,~A.~M.; Smith,~K.~A.; Kee,~R.~J. \emph{Electrochimica Acta}
  \textbf{2011}, \emph{58}, 33--43\relax
\mciteBstWouldAddEndPuncttrue
\mciteSetBstMidEndSepPunct{\mcitedefaultmidpunct}
{\mcitedefaultendpunct}{\mcitedefaultseppunct}\relax
\EndOfBibitem
\bibitem[R{\"o}der \latin{et~al.}(2016)R{\"o}der, Braatz, and
  Krewer]{roder2016multi}
R{\"o}der,~F.; Braatz,~R.; Krewer,~U. \emph{Computer Aided Chemical
  Engineering} \textbf{2016}, \emph{38}, 157\relax
\mciteBstWouldAddEndPuncttrue
\mciteSetBstMidEndSepPunct{\mcitedefaultmidpunct}
{\mcitedefaultendpunct}{\mcitedefaultseppunct}\relax
\EndOfBibitem
\bibitem[Hao \latin{et~al.}(2017)Hao, Liu, Balbuena, and
  Mukherjee]{hao2017mesoscale}
Hao,~F.; Liu,~Z.; Balbuena,~P.~B.; Mukherjee,~P.~P. \emph{The Journal of
  Physical Chemistry C} \textbf{2017}, \emph{121}, 26233--26240\relax
\mciteBstWouldAddEndPuncttrue
\mciteSetBstMidEndSepPunct{\mcitedefaultmidpunct}
{\mcitedefaultendpunct}{\mcitedefaultseppunct}\relax
\EndOfBibitem
\bibitem[Shi \latin{et~al.}(2012)Shi, Lu, Liu, Qi, Hector, Li, and
  Harris]{Shi2012}
Shi,~S.; Lu,~P.; Liu,~Z.; Qi,~Y.; Hector,~L.~G.; Li,~H.; Harris,~S.~J.
  \emph{Journal of the American Chemical Society} \textbf{2012}, \emph{134},
  15476--15487\relax
\mciteBstWouldAddEndPuncttrue
\mciteSetBstMidEndSepPunct{\mcitedefaultmidpunct}
{\mcitedefaultendpunct}{\mcitedefaultseppunct}\relax
\EndOfBibitem
\bibitem[Smith \latin{et~al.}(2011)Smith, Burns, Zhao, Xiong, and
  Dahn]{Smith2011}
Smith,~a.~J.; Burns,~J.~C.; Zhao,~X.; Xiong,~D.; Dahn,~J.~R. \emph{Journal of
  The Electrochemical Society} \textbf{2011}, \emph{158}, S23\relax
\mciteBstWouldAddEndPuncttrue
\mciteSetBstMidEndSepPunct{\mcitedefaultmidpunct}
{\mcitedefaultendpunct}{\mcitedefaultseppunct}\relax
\EndOfBibitem
\bibitem[Steinhauer \latin{et~al.}(2017)Steinhauer, Stich, Kurniawan,
  Seidlhofer, Trapp, Bund, Wagner, and Friedrich]{steinhauer2017situ}
Steinhauer,~M.; Stich,~M.; Kurniawan,~M.; Seidlhofer,~B.-K.; Trapp,~M.;
  Bund,~A.; Wagner,~N.; Friedrich,~K.~A. \emph{ACS Applied Materials \&
  Interfaces} \textbf{2017}, \emph{9}, 35794--35801\relax
\mciteBstWouldAddEndPuncttrue
\mciteSetBstMidEndSepPunct{\mcitedefaultmidpunct}
{\mcitedefaultendpunct}{\mcitedefaultseppunct}\relax
\EndOfBibitem
\bibitem[Deshpande \latin{et~al.}(2015)Deshpande, Ridgway, Fu, Zhang, Cai, and
  Battaglia]{deshpande2015limited}
Deshpande,~R.~D.; Ridgway,~P.; Fu,~Y.; Zhang,~W.; Cai,~J.; Battaglia,~V.
  \emph{Journal of The Electrochemical Society} \textbf{2015}, \emph{162},
  A330--A338\relax
\mciteBstWouldAddEndPuncttrue
\mciteSetBstMidEndSepPunct{\mcitedefaultmidpunct}
{\mcitedefaultendpunct}{\mcitedefaultseppunct}\relax
\EndOfBibitem
\bibitem[Borodin \latin{et~al.}(2015)Borodin, Olguin, Spear, Leiter, and
  Knap]{Borodin2015}
Borodin,~O.; Olguin,~M.; Spear,~C.~E.; Leiter,~K.~W.; Knap,~J.
  \emph{Nanotechnology} \textbf{2015}, \emph{26}, 354003\relax
\mciteBstWouldAddEndPuncttrue
\mciteSetBstMidEndSepPunct{\mcitedefaultmidpunct}
{\mcitedefaultendpunct}{\mcitedefaultseppunct}\relax
\EndOfBibitem
\bibitem[Latz and Zausch(2011)Latz, and Zausch]{Latz2011}
Latz,~A.; Zausch,~J. \emph{Journal of Power Sources} \textbf{2011}, \emph{196},
  3296--3302\relax
\mciteBstWouldAddEndPuncttrue
\mciteSetBstMidEndSepPunct{\mcitedefaultmidpunct}
{\mcitedefaultendpunct}{\mcitedefaultseppunct}\relax
\EndOfBibitem
\bibitem[Latz and Zausch(2013)Latz, and Zausch]{Latz2013}
Latz,~A.; Zausch,~J. \emph{Electrochimica Acta} \textbf{2013}, \emph{110},
  358--362\relax
\mciteBstWouldAddEndPuncttrue
\mciteSetBstMidEndSepPunct{\mcitedefaultmidpunct}
{\mcitedefaultendpunct}{\mcitedefaultseppunct}\relax
\EndOfBibitem
\bibitem[Borodin \latin{et~al.}(2006)Borodin, Smith, and
  Fan]{borodin2006molecular}
Borodin,~O.; Smith,~G.~D.; Fan,~P. \emph{The Journal of Physical Chemistry B}
  \textbf{2006}, \emph{110}, 22773--22779\relax
\mciteBstWouldAddEndPuncttrue
\mciteSetBstMidEndSepPunct{\mcitedefaultmidpunct}
{\mcitedefaultendpunct}{\mcitedefaultseppunct}\relax
\EndOfBibitem
\bibitem[Zhuang \latin{et~al.}(2005)Zhuang, Xu, Yang, Jow, and
  Ross]{zhuang2005lithium}
Zhuang,~G.~V.; Xu,~K.; Yang,~H.; Jow,~T.~R.; Ross,~P.~N. \emph{The Journal of
  Physical Chemistry B} \textbf{2005}, \emph{109}, 17567--17573\relax
\mciteBstWouldAddEndPuncttrue
\mciteSetBstMidEndSepPunct{\mcitedefaultmidpunct}
{\mcitedefaultendpunct}{\mcitedefaultseppunct}\relax
\EndOfBibitem
\bibitem[Edstr{\"o}m \latin{et~al.}(2006)Edstr{\"o}m, Herstedt, and
  Abraham]{edstrom2006new}
Edstr{\"o}m,~K.; Herstedt,~M.; Abraham,~D.~P. \emph{Journal of Power Sources}
  \textbf{2006}, \emph{153}, 380--384\relax
\mciteBstWouldAddEndPuncttrue
\mciteSetBstMidEndSepPunct{\mcitedefaultmidpunct}
{\mcitedefaultendpunct}{\mcitedefaultseppunct}\relax
\EndOfBibitem
\bibitem[Kulova \latin{et~al.}(2006)Kulova, Skundin, Nizhnikovskii, and
  Fesenko]{kulova2006temperature}
Kulova,~T.; Skundin,~A.; Nizhnikovskii,~E.; Fesenko,~A. \emph{Russian Journal
  of Electrochemistry} \textbf{2006}, \emph{42}, 259--262\relax
\mciteBstWouldAddEndPuncttrue
\mciteSetBstMidEndSepPunct{\mcitedefaultmidpunct}
{\mcitedefaultendpunct}{\mcitedefaultseppunct}\relax
\EndOfBibitem
\end{mcitethebibliography}

\end{document}